\documentclass[twocolumn,prl,superscriptaddress,floatfix,showpacs]{revtex4-1}
\usepackage[utf8]{inputenc}
\usepackage{amsmath}
\usepackage{amssymb}
\usepackage{graphicx}

\usepackage{graphics}
\usepackage{epsfig}

\newcommand{\be}{\begin{equation}} \newcommand{\ee}{\end{equation}}
\newcommand{\bea}{\begin{eqnarray}} \newcommand{\eea}{\end{eqnarray}}

\begin{document}

\title{Universality of critically pinned interfaces in 2-dimensional isotropic random media}

\author{Peter Grassberger}

\affiliation{JSC, FZ J\"ulich, D-52425 J\"ulich, Germany}

\date{\today}
\begin{abstract}
Based on extensive simulations, we conjecture that critically pinned interfaces in 2-dimensional 
isotropic random media with short range correlations are always in the universality class of 
ordinary percolation. Thus, in contrast to interfaces in $>2$ dimensions, there is no 
distinction between fractal (i.e., percolative) and rough but non-fractal interfaces. Our 
claim includes interfaces in zero-temperature random field Ising models (both with and without 
spontaneous nucleation), in heterogeneous 
bootstrap percolation, and in susceptible-weakened-infected-removed (SWIR) epidemics. It 
does not include models with long range correlations in the randomness, and models where 
overhangs are explicitly forbidden (which would imply non-isotropy of the medium). 
%We also discuss {\it nearly} pinned facetted interfaces, which occur in velocity
%controlled (``invasion") instead of force controlled (``epidemic") models, 
%in parameter regions where the latter would not give any growth at all.
\end{abstract}
\maketitle

Interfaces between different phases are important in many fields of physics, material sciences,
and biology. One might just mention boundaries of magnetic domains, cell membranes, or combustion
fronts. Because of that, there exists a huge literature \cite{family,barabasi}. Maybe best 
understood are moving interfaces (the most famous model there being that of Kardar, Paris, and Zhang
\cite{KPZ}), but in the present paper we shall deal with interfaces that, although being pushed,
get pinned at random obstacles. More precisely, we are interested in {\it critically} pinned
interfaces, where the pushing force is just such that the movement gets slower and slower, and 
would stop at infinite time. We assume that the medium in which the interface moves is isotropic
with short range correlations, and we neglect thermal fluctuations. Finally, we will only be 
interested in 2 dimensions, where the interface is a line.

Since the literature on pinned (and moving) interfaces is dominated by models where overhangs 
are neglected \cite{Nattermann,Giamarchi,LeDoussal}, 
we should stress that such models do {\it not}, in principle, describe isotropic
media. It {\it may} be that overhangs turn out to be unimportant on large scales, but that is 
something that should not be imposed from the very beginning. There is a consensus that they 
are indeed irrelevant for {\it growing} interfaces, but 
the situation is much less clear for pinned interfaces -- and, in particular, in two dimensions. 
Therefore, overhangs will in the following be fully taken into account.

Maybe the oldest and best understood model of such interfaces (although it is often not considered 
as such) is percolation \cite{stauffer}. The reason why boundaries of critical percolation clusters
are usually not considered as interfaces is that these clusters are fractal, and therefore cannot 
be considered as a (bulk) phase. But this does not take into account the possibility that maybe 
there are no critically pinned interfaces at all that separate two non-fractal phases, i.e. that 
interfaces maybe {\it always} are between fractal phases.

It is well known \cite{Ji,Koiller1,Koiller2,Koiller3} that 
2-d interfaces in 3-d random media can be either fractal (as in ordinary percolation, OP) or rough 
but non-fractal. Indeed, it was shown by Janssen {\it et al.} \cite{Janssen04}, that these two
cases can be realized in the same model, and are separated by a tricritical point, in any dimension 
$\geq 3$. This was verified numerically in \cite{Bizhani}, where the tricritical exponents were 
measured precisely in $d=3$, and where it was found that no such tricritical point exists for $d=2$.

If the latter is correct and general, this would mean that critically pinned interfaces in 2-d
isotropic media  are {\it always} in the universality class of OP. This had been 
conjectured before \cite{Drossel}, but the opposite was claimed in many, even very recent, papers 
\cite{Cieplak1,Cieplak2,Martys1,Martys2,Nolle,Frontera-1999,Zhou-2010,Qin-2012,Xi-2015,Spasojevic,Janicevic-2017,Thongjaomayum,Kurbah}
In most of these papers \cite{Cieplak1,Cieplak2,Martys1,Martys2,Nolle,Frontera-1999,Seppala,Thongjaomayum,Kurbah} 
it is claimed that the percolation scenario breaks down when the disorder is weak. 
But in others \cite{Xi-2015,Spasojevic,Janicevic-2017}, even
in the percolation like phase the critical exponents were found to be different. In \cite{Qin-2012}, 
even continuously varying exponents were found. In \cite{Seppala}, it was correctly claimed that 
the fractal-to-rough transitions seen in previous papers were finite size effects and that interfaces 
in the random field Ising model (RFIM) in $d=2$ are in the percolation universality class, but it was 
also claimed that the critical field strength for small disorder is zero, which seems to contradict
our results. Finally, in \cite{Si-2016} exponents were found that 
are in rough agreement with those for OP, but no connection was suggested.

Indeed, it is known \cite{Aizenman} 
that no phase transitions can exist in Ising-type models with quenched disorder, even at zero 
temperature. In principle, this should exclude any tricritical point as found by Janssen et al. in 
higher dimensions, but the situation is somewhat more subtle \cite{Balog}, since the existence of 
non-fractal pinned interfaces does not necessarily imply that the two phases separated by them are 
{\it thermodynamically} stable.

It is the purpose of the present paper to clarify this situation, by means of simulations that 
show clearly that the transitions proposed in earlier papers were just cross-overs that exist only 
for finite systems, and that the exponents are always those of OP. 

The model used in our simulations is the generalization of site and bond percolation introduced in 
\cite{Bizhani}. In this model, the probability that a site surrounded by $n$ ``wetted" (infected, flipped)
sites will ultimately also be wetted is $q_n$. Put differently, if these neighbors ``attack" the site
in any sequential order, then the probability for the site to fall during the $m-$th attack is $p_m$
with 
\be
    q_n = q_{n-1}+(1-q_{n-1})p_n:
\ee
if the site has fallen after $n$ attacks, then either it had already fallen before the last one, or 
(if not) it falls during the last. Site percolation corresponds to $p_1=p, p_m=0$ for $m>1$. Bond 
percolation has $p_m = p$ for all $m$.

The simplest non-trivial model in this class, called the ``minimal model" (MM) in the following, has
$p_m=p_2$ for all $m>2$. In this model one just has to distinguish between first and 
subsequent attacks, but not between second, third, etc. This is also identical to the SWIR model of 
\cite{Chung-2014,Chung-2016,Lee-2017,Choi-2017} for the special case where sites get weakened with 
probability 1 after an attack. The general SWIR model with finite weakening probability has non-trivial 
$p_m$'s that can easily obtained recursively \cite{SM}. Finally, also the heterogeneous $k$-core 
percolation model of \cite{Doro,Cellai} is equivalent to the above model with general $p_m$'s. This is 
less obvious, because in \cite{Doro,Cellai} sites do not fall with ad hoc determined probabilities, but have
probabilities to be in the $k$-core that were chosen before any spreading. But since each site can get
infected at most once, it is easy to see that both points of view are equivalent.

Finally, also the RFIM model is a special case of the above \cite{Drossel}. Consider the Hamiltonian
$H = -J\sum{\langle i,j\rangle} S_i S_j - \sum_i (h_i + H)S_i$ with $S_i = \pm 1$, where we can take 
$J=1$ without loss of generality, and where the local fields $h_i$ are distributed according to some 
probability $P(h)$. In the following, we shall consider explicitly the uniform distribution 
$P(h) = \Theta(\Delta-|h|)/2\Delta$, and the Gaussian distribution 
$P(h) = 1/(\sqrt{2\pi}\sigma) \exp[h^2/(2\sigma^2)]$.

Consider the case where we start with a very strong negative external field, $H \ll 0$, so that all 
spins are initially $S_i=-1$. Then we increase $H$ continuously. A spin $i$ surrounded by $n$ ``up"
and ${\cal N}$ ``down" spins (${\cal N}$ is the coordination number; on the square lattice, ${\cal N}=4$) 
will flip, if this becomes energetically favorable, $H > {\cal N}-2n - h_i$. Therefore,
\be
    q_n = \int_{{\cal N}-2n - H}^\infty P(h) dh.    \label{q_RFIM}
\ee
The final expressions for uniform and Gaussian distributions are given in \cite{SM}.

Notice that this gives in general also $q_0>0$, i.e. spins can also flip spontaneously without any
flipped neighbors, i.e. clusters can nucleate \cite{foot1}. We have then two options: we can either 
follow the bulk of the literature on the kinetic zero-temperature RFIM and forbid spontaneous
flips {\it by fiat}, or we can allow them (we do not allow spontaneous flips of multiple spins).
In the former case we have to start with a {\it seed} (we take wither a single point seed or an entire
line on which the spins are already flipped initially). In the latter case, we can use the same code,
provided we include in the seed those sites that would flip spontaneously. The actual codes used are 
described in \cite{SM}.

\begin{figure}
\begin{centering}
\vglue -1.1cm
\includegraphics[scale=0.36]{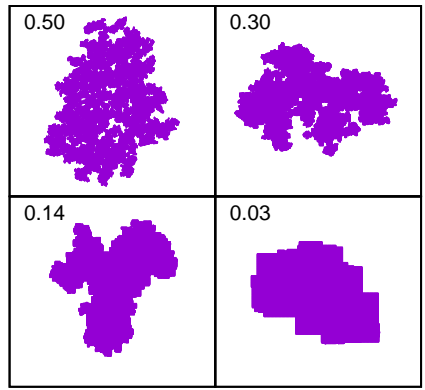}
\par\end{centering}
\vglue -0.7cm
\caption{\label{fig1} (color online)  Typical critical clusters obtained from single site seeds with 
   the MM. The numbers written in each panel show the values of $p_1$. The values of $p_2$ are 
   $0.50, 0.88092, 0.977452$, and $0.99897$. The upper left panel corresponds to bond percolation.}
\end{figure}

Typical critical clusters obtained with the MM are shown in Fig.~1. While the clusters in the upper 
row are obviously fractal (the left one is indeed bond percolation), the ones in the lower row 
{\it seem} to be compact. It is this which has misled many authors to claim that the growth is indeed 
compact when $p_1$ is small. The reason why the clusters look less fractal for small $p_1$ is clear:
small $a_1p$ and large $p_2$ mean that growth into new territory is slowed down, while bays and fjords are 
filled up. 

\begin{figure}
\begin{centering}
\includegraphics[scale=0.30]{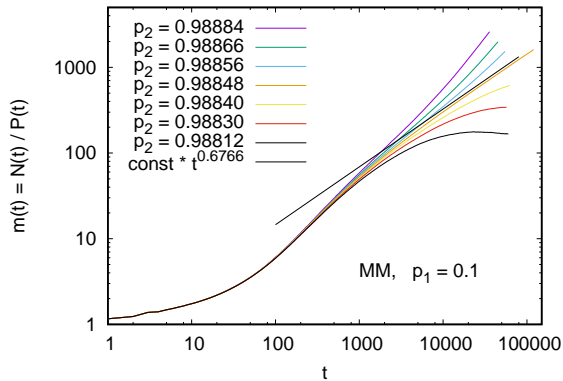}
\par\end{centering}
\vglue -0.5cm
\caption{\label{fig2} (color online) Log-log plot of the average number of growth sites per surviving 
   cluster in the MM, plotted 
   against time $t$. Each curve corresponds to a different value of $p_2$, while $p_1=0.1$ is common 
   to all curves. The straight line has the slope expected for OP.}
\end{figure}

\begin{figure}
\begin{centering}
\includegraphics[scale=0.30]{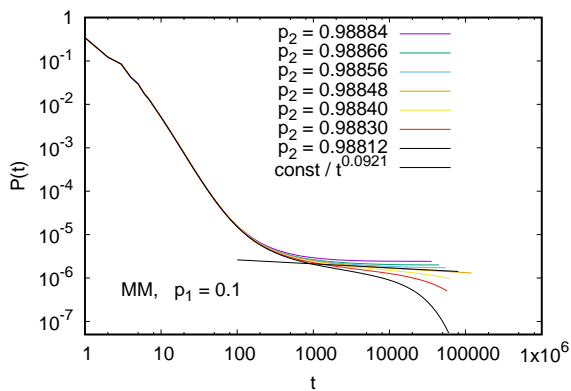}
\vglue -0.5cm
\includegraphics[scale=0.30]{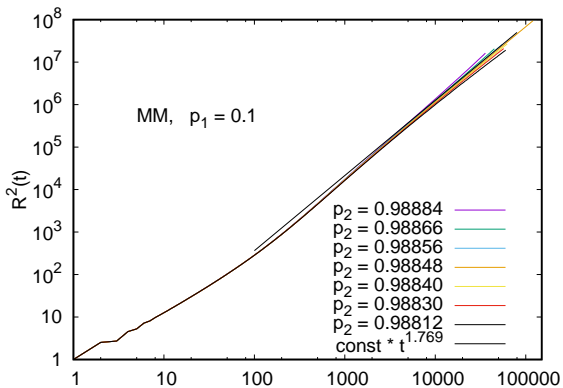}
\vglue -0.5cm
\par\end{centering}
\caption{\label{fig3} (color online) Similar to Fig.~2, but for the survival probability $P(t)$ (top panel)
   and for $R^2(t)$ (bottom panel).}
\end{figure}

The fact that this seeming compactness is only transient is best seen by studying the 
usual observables for percolation growth: The number $N(t)$ of growth sites at time $t$, the survival
probability $P(t)$, and the r.m.s. distance $R^2(t)$ of the growth sites from the seed. The scaling 
laws for OP are summarized in \cite{SM}. 

In Fig.~2 we show $N(t)/P(t)$ for $p_1=0.1$ 
and for several values of $p_2$ close to $p_{2,c}$. We see the expected huge deviations from scaling for 
small $t$, but for $t>10^4$ there is one straight curve, and it has precisely the exponent of the percolation
universality class. This might be an accident, but as seen from Fig.~3, the same value of $p_2$ 
gives also power laws for $P(t)$ and $R^2(t)$ with exactly the right critical exponents. Blow-ups of 
these figures, where we have also divided the data by the expected power laws to obtain horizontal curves 
at criticality, are given in \cite{SM}.

\begin{figure}
\begin{centering}
\includegraphics[scale=0.30]{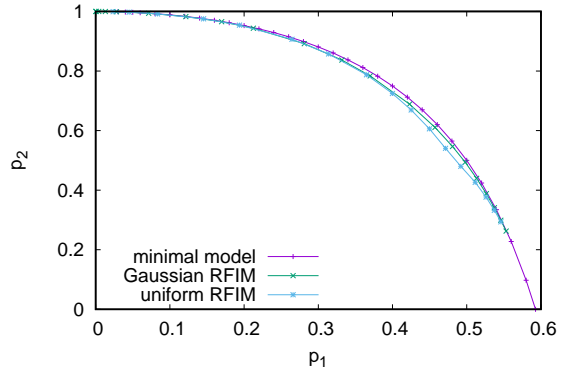}
\par\end{centering}
\caption{\label{fig4} (color online) Phase boundaries for the minimal model and for two versions of the 
   RFIM. Interfaces are moving in the region above the curves, and pinned below. Error bars on all points
   are much smaller than the symbols.}
\end{figure}

\begin{figure}
\begin{centering}
\includegraphics[scale=0.30]{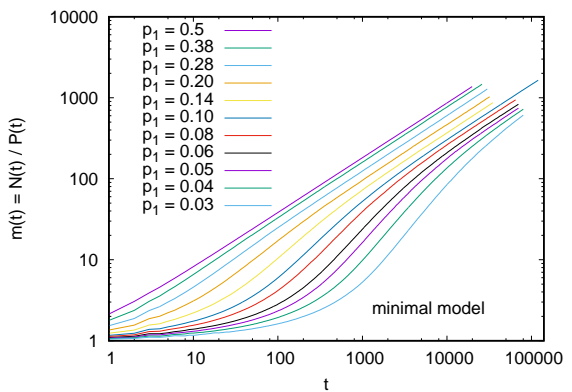}
\par\end{centering}
\caption{\label{fig5} (color online) Log-log plots of $m(t)=N(t)/P(t)$ for $p_2=p_{2,c}(p_1)$, i.e. on the 
   critical curve. For small $p_1$, clusters hardly grow at the beginning, but the long time growth 
   is that of OP.}
\end{figure}

Similar analyses were also made for 30 other values of $p_1$, in the range $0.03 \leq p_1 \leq 0.592746$. For 
each of them we obtained $q_c$ with errors $\approx 10^{-5}$, see Fig.~4 (remember that $p=p_1$ and $q=p_2$
for the MM). Some of the critical curves of $N(t)/P(t)$ versus $t$ are shown in Fig.~5. They show 
increasingly large deviations from scaling as $p\to 0$, but they all show the scaling of ordinary 
percolation for large $t$. The deviations at small $t$ arise from the fact that only very clusters
survive the initial growth phase, and even if they do they grow very slowly. In the limit $p\to 0$, 
each surviving cluster is just a single site that performs a self-avoiding random walk. All this is 
clearly seen from plots analogous to Fig.~4, but for $P(t)$ and $R^2(t)$ \cite{SM}.

\begin{figure}
\begin{centering}
\includegraphics[scale=0.30]{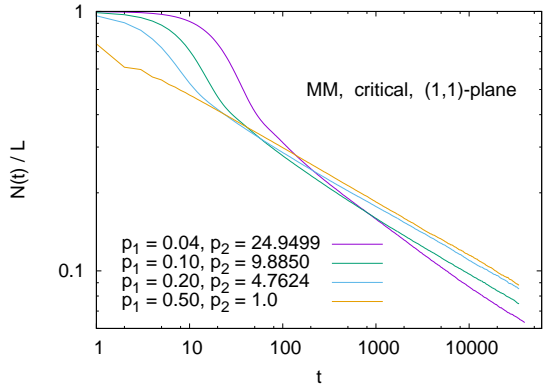}
\par\end{centering}
\caption{\label{fig6} (color online) Log-log plots of $N(t)/L$ for critical interfaces with
   global (1,1) orientation.}
\end{figure}

Let us now turn to line seeds, i.e. to initially flat interfaces. The decrease of the number of growth
sites with time is shown in Fig.~6, for critically pinned interfaces with (1,1) orientation. We see 
precisely the expected power laws \cite{SM} for large $t$, preceded for small $p$ by initial periods
where $N(t) \approx const.$ Data for interface heights are shown in \cite{SM}.

\begin{figure}
\begin{centering}
\includegraphics[scale=0.30]{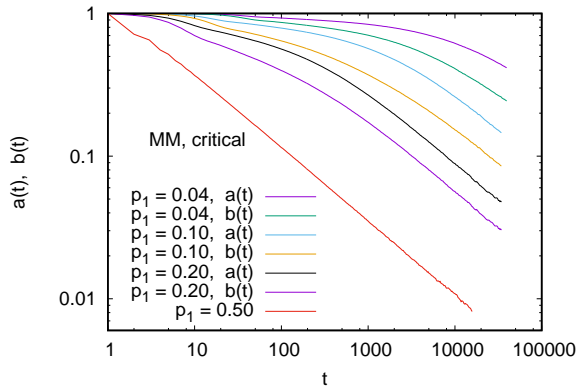}
\par\end{centering}
\caption{\label{fig7} (color online) Log-log plots of asymmetries $a(t)$ and $b(t)$ for critical
   interfaces with global (1,1) orientation. Notice that $a(t)=b(t)$ by definition for bond percolation.}
\end{figure}

For a last check that the MM is in the OP universality class, we studied the {\it local} 
orientation of globally flat interfaces. Local orientations are measured by the directions of either
{\it attacks}, i.e. any contacts between infected and susceptible sites, or {\it infections}, i.e. 
such contacts that lead actually to an infection of the immune site \cite{Janssen-contact,Grass-contact}.
Let us define $m_f$ and $m_b$ as the number of forward and backward attacks/infections, and 
\be
    a(t) = \frac{m_{f,\rm attack}-m_{b,\rm attack}}{m_{f,\rm attack}+m_{b,\rm attack}},\quad \\
    b(t) = \frac{m_{f,\rm infect}-m_{b,\rm infect}}{m_{f,\rm infect}+m_{b,\rm infect}}.
\ee
It was shown in \cite{Janssen-contact} that these should in general satisfy power laws with new
critical exponents, and it is shown in \cite{Grass-contact} that $a(t) = b(t) \sim t^{-0.5189(1)}$ 
for OP. Data for $a(t)$ and $b(t)$ for the MM are shown in Fig.~7, where all curves 
show indeed for large $t$ the expected scaling.

The discussion of the RFIM can now be very short: For all observables, we found very similar behavior,
to the point that showing these results is hardly of any use in most cases (some data are shown in 
\cite{SM}). We just have to remember that we can, for any $H$ and any disorder strength, determine 
the corresponding $p_m$. Critical values of $p_1$ and $p_2$ for the case without spontaneous nucleation
are shown in Fig.~4 together with the values for the MM. They are 
very similar. Both RFI models converge to site percolation ($p_1=0.5927\ldots, p_2 =0$) along 
the same curve for strong disorder, and behave exactly like the MM for weak disorder (where $q_n=1$
for $n>2$).

\begin{figure}
\begin{centering}
\includegraphics[scale=0.30]{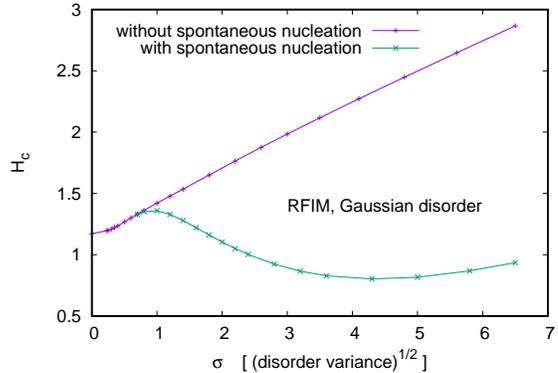}
\par\end{centering}
\caption{\label{fig8} (color online) Depinning thresholds for the RFIM with Gaussian disorder. The value 
   for $\sigma\to 0$ is an exact analytic result \cite{Drossel}, the other points are from simulations 
   and have error bars much smaller than the symbols.}
\end{figure}

The only model demanding more discussions is the RFIM with spontaneous nucleation. For weak uniform
disorder and not too large $H$, single sites cannot flip spontaneously. In that case, the simulations
where nucleation is forbidden represent the true non-equilibrium RFIM. This is not so for Gaussian
disorder, where spontaneous single spin flips can always occur. They are very rare for small 
disorder, but they become important and change the phase boundary completely for large disorder, 
see Fig.~8. The most striking aspect of this plot is the huge difference with the results of 
\cite{Seppala}, who found much smaller values of $H_c$, and conjectured $H_c=0$ for $\sigma < 0.6$.
Presumably this is due to the fact that we allowed only single spin spontaneous flips, while the 
states in \cite{Seppala} were stable against flips of {\it any} finite clusters. We checked that 
flipping clusters with two-site seeds did not change our results, but flipping larger clusters with
larger seeds would have been impossible with our algorithm.

Let us finally comment on facetted growth, as seen e.g. in
\cite{Ji,Koiller1,Cieplak1,Cieplak2,Martys1,Martys2}. Obviously, this cannot occur in strictly 
isotropic media, since it requires a regular lattice for the orientations of the facets (notice 
that we also discussed lattice models in this paper, but Fig.~7 showed that indeed the lattice
anisotropy became irrelevant in the scaling limit). In addition, it seems that facetted growth 
never is critical. In \cite{SM} we discuss this for the RFIM with uniform disorder, where a 
first order transition was claimed in \cite{Qin-2012,Si-2016}.

In summary, we found numerically that in all models the depinning transition is in the ordinary 
percolation universality class. In \cite{Cai,Grass-coinfect} it was also found that interfaces 
in a 2-d SIR type model of coinfections are in the OP class (in contrast to higher dimensions).
Together with analytical arguments this gives strong support to our more general claim. It also 
suggests that a percolative phase transition in a 2-d model with intermediate length dependency 
links \cite{Li-Bashan} is continuous and in the OP class, as claimed in\cite{Grass-SOC} and in 
contrast to claims made in \cite{Li-Bashan}.

Acknowledgements: I thank Golnoosh Bizhani and Maya Paczuski for collaborations during very 
early stages of this work.


\begin{thebibliography}{10}
\bibitem{foot1} For Gaussian disorder with any $\sigma$ this can happen for any $H$, but for uniform disorder 
   single spins can flip spontaneously only if $H+\Delta >{\cal N}$.
\end{thebibliography}
\end{document}

\title{Supplemental Material:\\
Universality of critically pinned interfaces in 2-dimensional isotropic random media}

\maketitle

\section{The SWIR model}

The Susceptible-Weakened-Infected-Removed (SWIR) model was defined in \cite{Chung-2014} as a 
generalization of the standard Susceptible-Infected-Removed (SIR) model, in which susceptible
sites can become weakened with probability $\mu$ by infected neighbors, while they get infected
with probability $\kappa$. Upon further contacts, weakened sites become infected with probability 
$\eta$. Finally, infected sites are removed with probability one.

Let us denote by $p_m(S), p_m(W),p_m(I),$ and $p_m(R)$ the probabilities that an originally 
susceptible site is in state $S,W,I,$ and $R$ after $m$ contacts. Then we have the recursion
\bea
   p_{m+1}(S) & = & (1-\kappa-\mu) p_m(S) \nonumber \\ 
   p_{m+1}(W) & = & \mu p_m(S) + (1-\eta) p_m(W) \nonumber \\
   p_{m+1}(I) & = & \kappa p_m(S) + \eta p_m(W) \nonumber \\
   p_{m+1}(R) & = & p_{m}(I),
\eea
and finally $q_n = p_n(R)$.

\section{The RFIM}

In the main text we have seen that the RFIM is a special case of the model of \cite{Bizhani},
with probabilities $q_n$ given by Eq.~(2). For the uniform and Gaussian distributions on the 
square lattice this gives
\be
   q_{n,\rm uniform} = (\Delta+H+2n-4)/\Delta
\ee
and
\be
   q_{n,\rm gauss} = \frac{1}{2} {\rm erfc}\left(\frac{4-2n-H}{\sqrt{2}\sigma}\right).
\ee

\section{Simulations}

All simulations described in the paper were done with straightforward generalizations of the 
classical Leath \cite{leath} algorithm. In this algorithm one has three data structures: 

\begin{itemize}

\item A byte map, where for every site its status is stored. In the MM, we have to store for each 
site only whether it was never attacked, whether it had been attacked, or whether it is removed.
For the general SWIR model and for both versions of the RFIM, we have to store also the number of previous
attacks. Alternatively, we can also store for each site the number of future attacks at which the 
site will succumb. The latter of course requires that we determine these numbers in a first pass through 
the lattice. In contrast, if we store the numbers of past attacks, we can determine when the site
falls ``on the fly", which is faster if the growing cluster fills only a small part of the lattice.
On the other hand, for simulations of the RFIM with spontaneous nucleation, it is more efficient
(and simpler) to store the number of future attacks.

\item A list of ``active" sites. Initially, this contains all seed sites. In later steps (remember 
that time is discrete) it contains all sites ``wetted" (or flipped) in the last previous time step.

\item A list of ``growth sites". The growth sites at time $t$ are just those neighbors of the sites
that were active at time $t-1$, which actually got infected (wetted, flipped). When proceeding from 
time $t$ to $t+1$, the old list of active sites is replaced by the list of growth sites, and the 
new list of growth sites starts empty.
\end{itemize}

Indeed, this describes the ``breadth-first" implementation of the Leath algorithm, if we implement 
the two lists by arrays that are handled in first-in first-out style. Alternatively, we could have 
implemented them also ``depth-first", in which the lists are implemented by stacks handled in
first-in last-out style. 

While it is in general a matter of taste whether one prefers a breadth-first or a depth-first
implementation, the breadth-first one becomes essential for single site seeds and very small $q_0$.
In that case, most attempts to build a large cluster would fail, if the simple Leath algorithm were 
used, and we used instead a version of PERM (pruned-enhanced Rosenbluth method) \cite{PERM,animals}. 
This is based on re-sampling and which allows to grow clusters with good statistics, even if the 
probability for a cluster to grow is tiny. To be efficient, Leath growth with PERM has to be breadth
first \cite{animals}.

In simulations with single site seeds, we always verified that clusters never reached the lattice 
boundaries. In simulations with an entire line of seeds, we took this line as one of the four 
boundaries. Lateral boundary conditions were periodic, while the boundary opposite to the seed
was sufficiently far away that the clusters never reached them (this was also checked explicitly).
To minimize finite size corrections, we used for the runs with line seeds lattices with aspect ratios
$\geq 2$, i.e. the lattices were at least twice as long in the lateral direction than in the 
direction of growth.

Finally, in simulations of the RFIM with spontaneous nucleation, we have to add a last step in order 
to determine the interface. In these simulations we started with a seed which consists both of a (boundary)
line and of all interior sites that would flip spontaneously. After the Leath algorithm is run, the flipped
``cluster" consists both of the part that is attached to the line seed, and a large number of disconnected
small clusters. In order to remove the latter and to obtain the interface of the connected cluster
attached to the line seed, one has to run the Leath algorithm a second time, with only the line as a 
seed and where only sites are included in the growth that are already flipped after the main step.

Typical lattice sizes were $32678\times 32678$ for single site seeds, and 
$65536\times 16384$ or $65536\times 32678$ for line seeds. All simulations were done on work station
clusters and laptops. The total CPU time spent was about 2 years.

\section{Critical exponents for ordinary 2-d percolation}

The ``classical" exponents are $\beta=5/36$, $\nu=4/3$, and $D_f = 91/48$ \cite{Stauffer}, where $\beta$
controls the density of the infinite cluster in the slightly supercritical region, $\nu$ controls 
the divergence of the correlation length as one approaches the critical point, and $D_f$ is the 
fractal dimension at criticality. In order to compare with the simulations of {\it growing} clusters
as in the present paper, one also needs the minimal path exponent $d_{\rm min} = 1.130 77(2)$ \cite{Zhou},
controls the relation between Euclidean distances and ``chemical layers" or time. More precisely,
\be
    R^2(t) \sim t^{2/d_{\rm min}} \sim t^{1.76871(4)}.
\ee
exactly at the critical point. The survival probability $P(t)$ at the critical point scales as 
\be
    P(t) \sim t^{-\delta},\quad \delta = \beta/(\nu d_{\rm min}) =0.092120(2).
\ee
Finally, $m(t) = N(t)/P(t)$ is the average number of growth sites per surviving cluster. The sum
$M(T) = \sum_{t\leq T}$ is then ``mass" of a cluster grown for a time $T$ and thus having a radius
$R\sim T^{1/d_{\rm min}}$, and scales as $M \sim R^{D_f} \sim T^{D_f/d_{\rm min}}$, giving
\be
    m(t) \sim t^\mu,\quad \mu = D_f/d_{\rm min}-1 = 0.67659(3).
\ee

These are the scaling laws pertinent for single site seeds. For line seeds, the average distance of 
the growth sites from the seed also scales at criticality as $R\sim t^{1/d_{\rm min}}$, while the 
density of the final cluster decays as $\rho(R) \sim R^{D_f-2}$. Thus the number of growth sites on a 
lattice of lateral size $L$ should scale as 
\be
    N(t) \sim L t^{-\eta},\quad \eta =(D_f-1)/d_{\rm min}-1 = 0.20777(1).
\ee
The average height of the interface grows, finally, as $h(t)\sim t^{1/d_{\rm min}}$. This should be 
independent of the way how we define height. We can use, in particular, the average height of growth
sites or we can, for each lateral position $x$, define height as the maximal height of any infected
site.

\section{Point seed clusters for the MM}

\begin{figure}
\begin{centering}
\vglue -0.4cm
\includegraphics[scale=0.30]{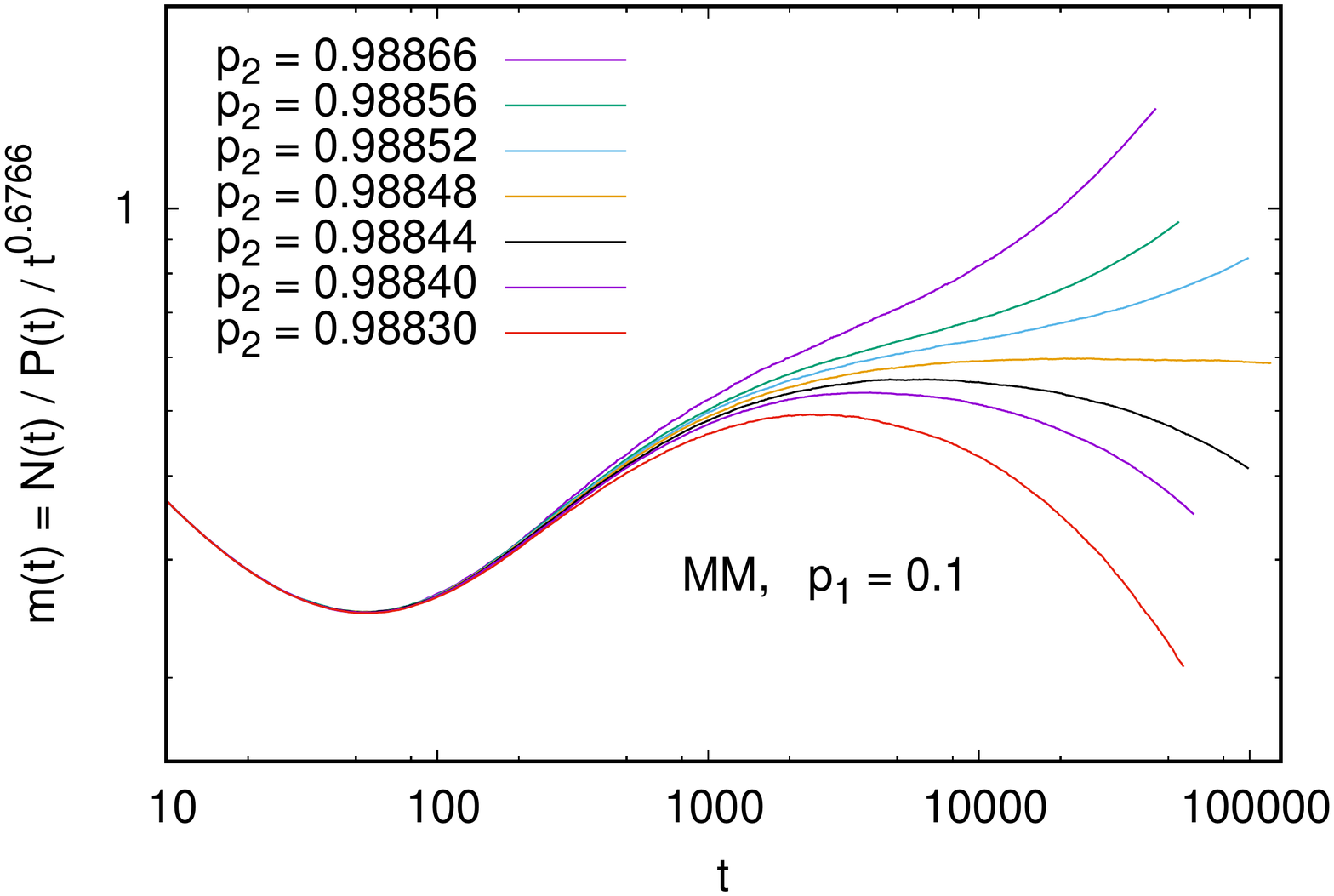}
\vglue -0.7cm
\includegraphics[scale=0.30]{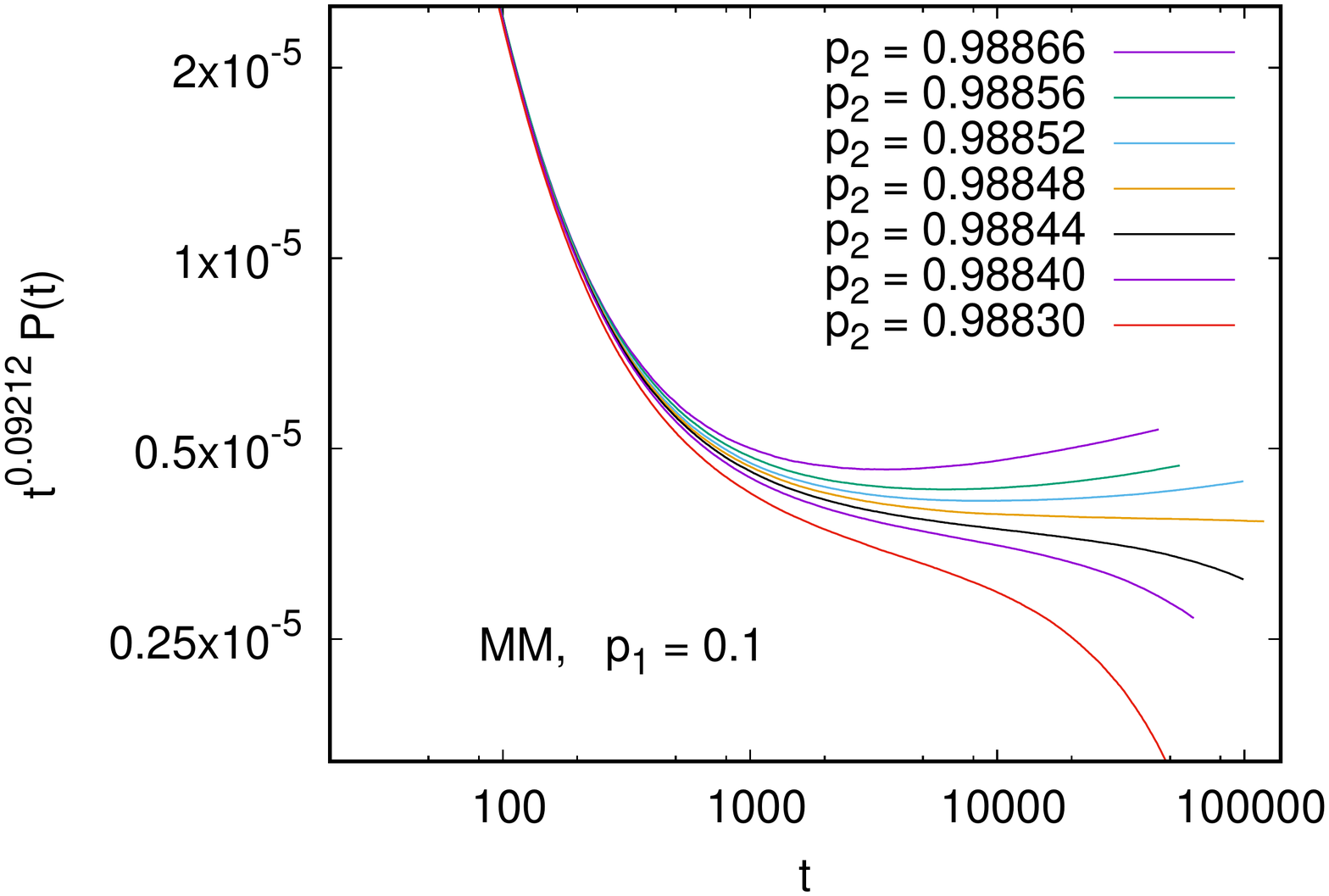}
\vglue -0.7cm
\includegraphics[scale=0.30]{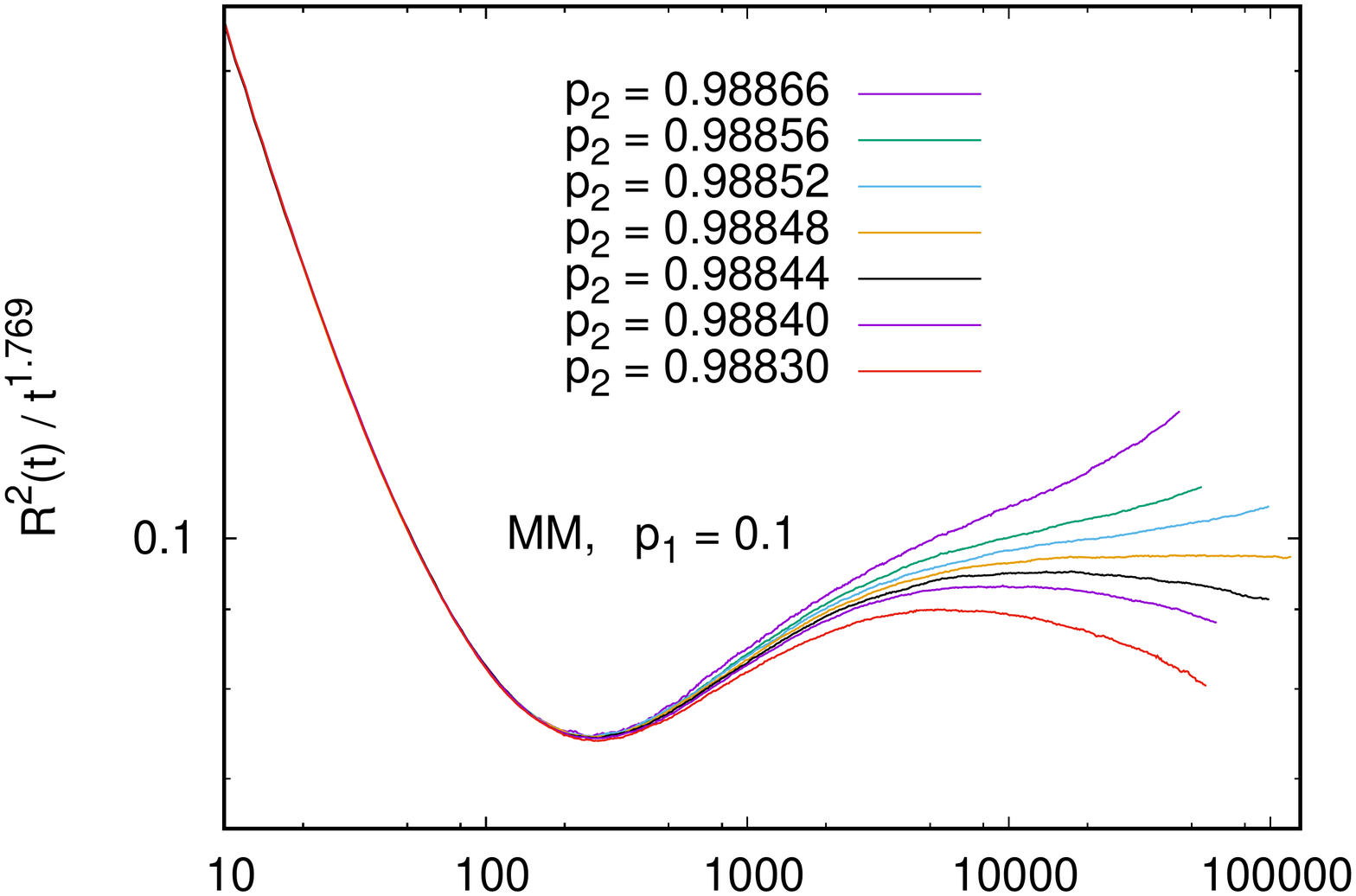}
\vglue -0.4cm
\par\end{centering}
\caption{\label{fig1} (color online) Same data as in Figs.~2 and 3 in the main text, but divided by 
   the expected power laws so that the critical curves become horizontal asymptotically.}
\end{figure}

\begin{figure}
\begin{centering}
\vglue -0.4cm
\includegraphics[scale=0.30]{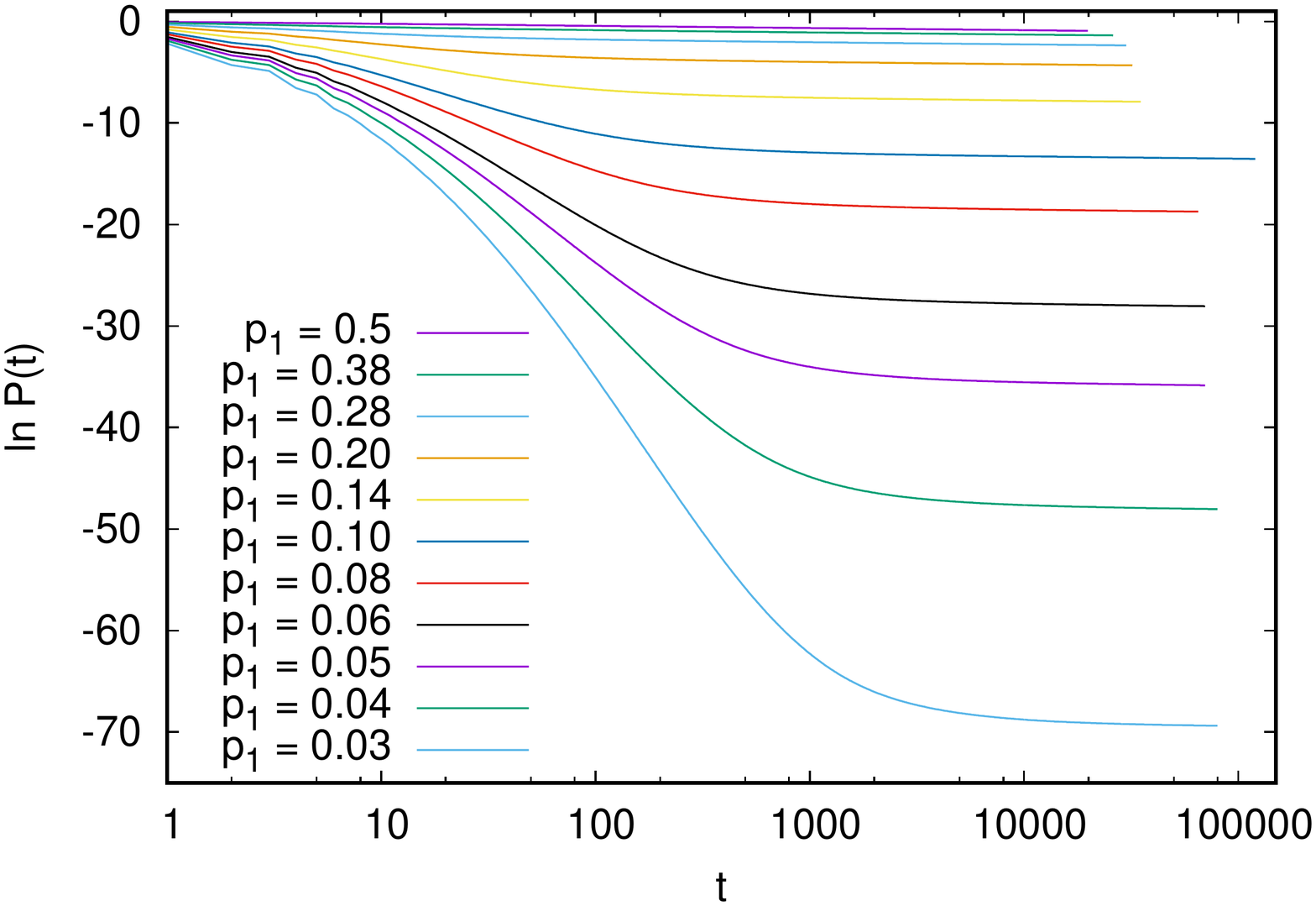}
\vglue -0.7cm
\includegraphics[scale=0.30]{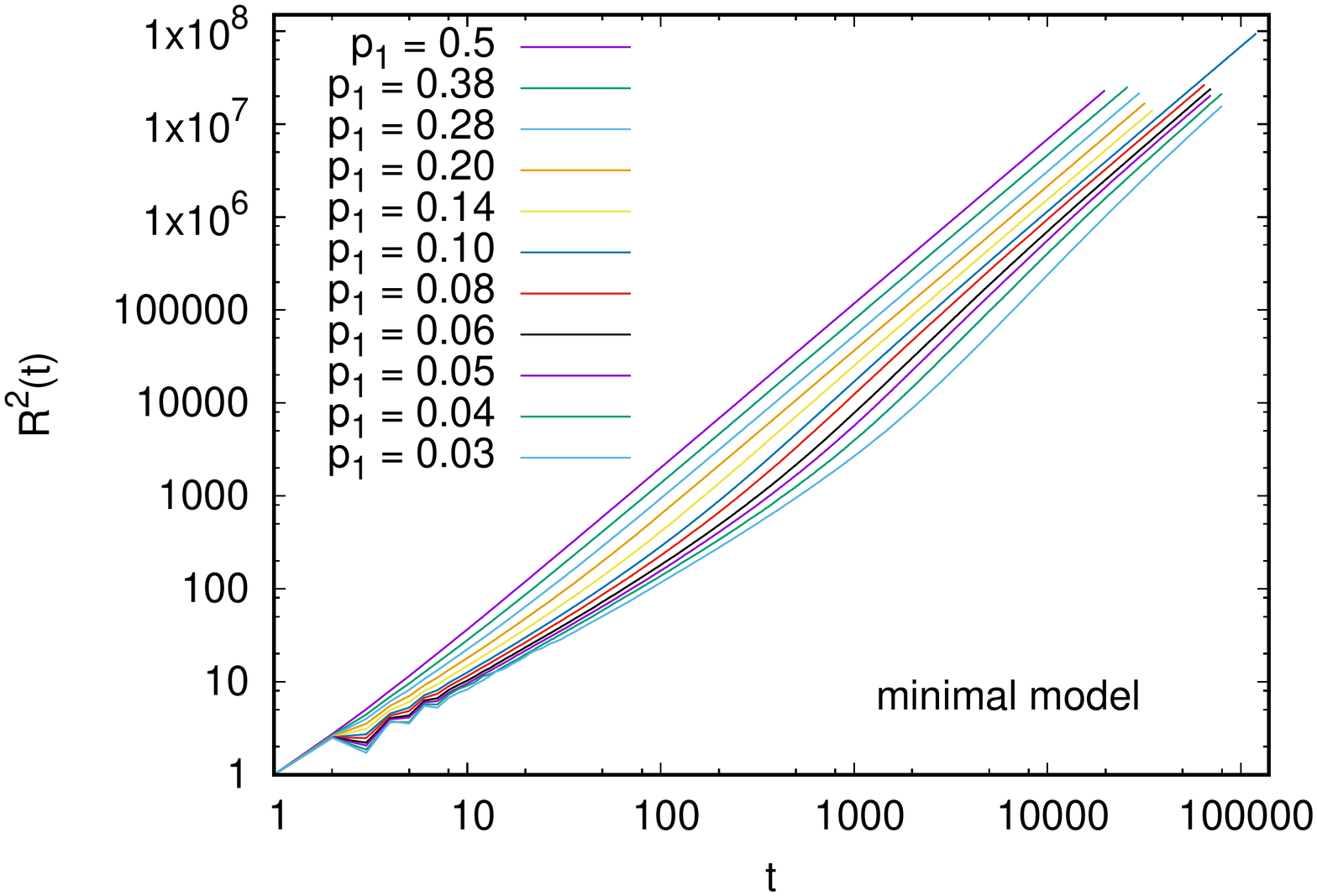}
\vglue -0.4cm
\par\end{centering}
\caption{\label{fig2} (color online) Plots of $\ln P(t)$ (upper panel) and $R^2(t)$ (lower panel) 
   against $t$, for the same 11 points on the critical pinning line for which $N(t)/P(t)$ is shown 
   in Fig.~5 of the main text.}
\end{figure}

In Fig.~S1 we show the large-$t$ data (obtained with PERM) shown already in Figs.~2 and 3 of the main
text, but multiplied with suitable power laws so that the critical curves are expected to be horizontal,
which allows much finer $y$-axes. We see that indeed the critical curves (corresponding to $p_1=0.1,
p_2 = 0.988482(2)$) are the most straight for large $t$ and have exactly the exponents expected for 
percolation.

\begin{figure}
\begin{centering}
\vglue -0.4cm
\includegraphics[scale=0.30]{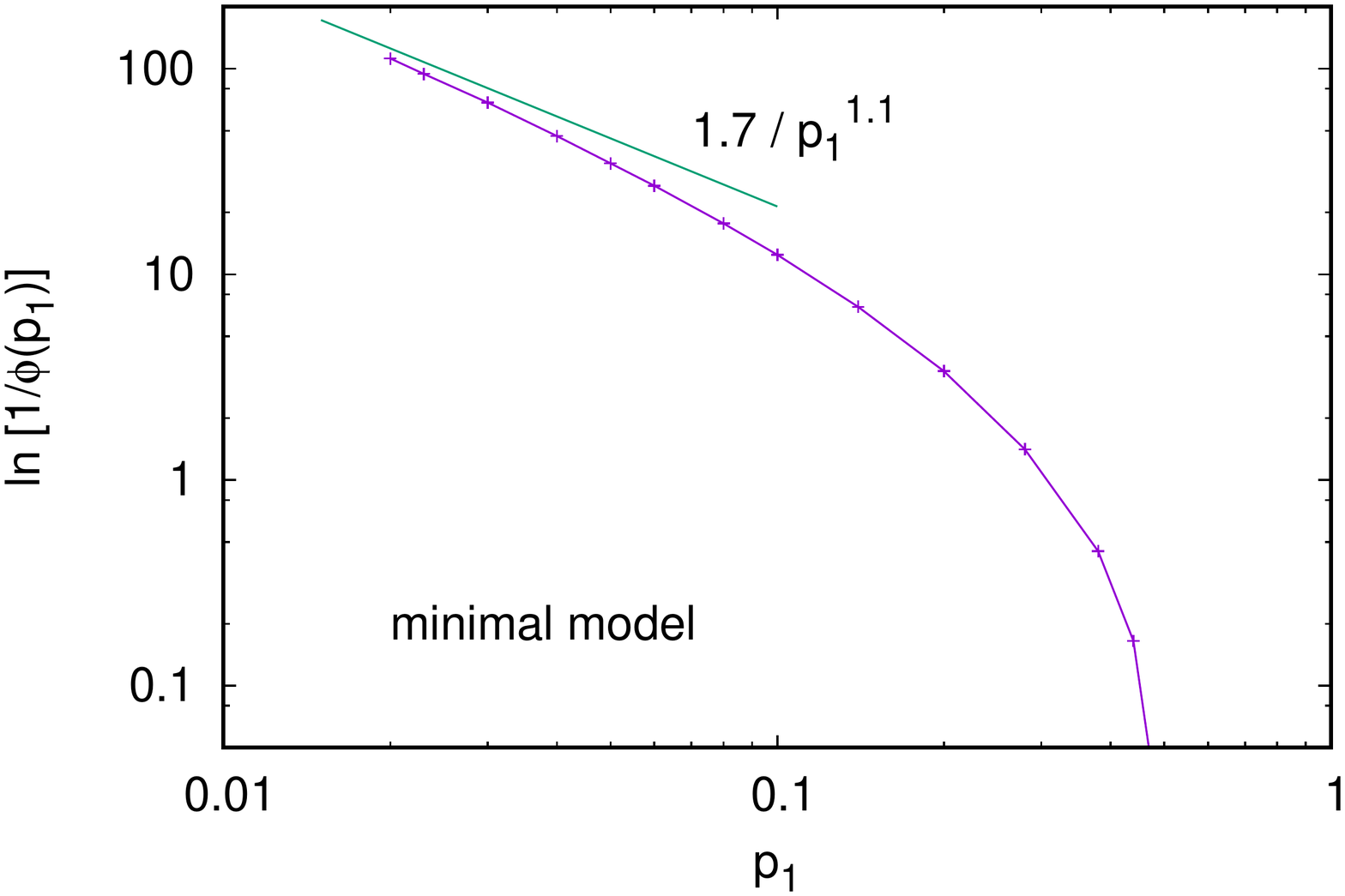}
\vglue -0.4cm
\par\end{centering}
\caption{\label{fig3} (color online) Log-log plot of $\ln [1/\phi(p)]$ versus $p$, where $\phi(p)$ is 
   the prefactor of the power law $P(t) \sim t^{-\delta}$. The straight line 
   corresponds to $\phi(p) = \exp(-1.7/p^{1.1})$.}
\end{figure}

In Fig.~S2 we show $P(t)$ (top panel) and $R^(t)$ (bottom panel) for the same 11 pairs $(p,q)$ as in 
Fig.~5 of the main text. Thus each curve corresponds to a point on the depinning line. We see that 
$P(t)$ tends to zero faster than any exponential as $p_1 \to 0$. The best fit for the large-$t$ behavior
is
\be
   P(t) \sim t^{-\delta} \phi(p)
\ee
with $\phi(p) \sim \exp({\rm -const}/p^{1.1})$ for $p\to 0$, see Fig.~S3. The short $t$ behavior of 
$R^2(t)$ is for $p\to 0$ compatible with a self avoiding random walk of a single-site ``cluster".

\begin{figure}
\begin{centering}
\vglue -0.4cm
\includegraphics[scale=0.30]{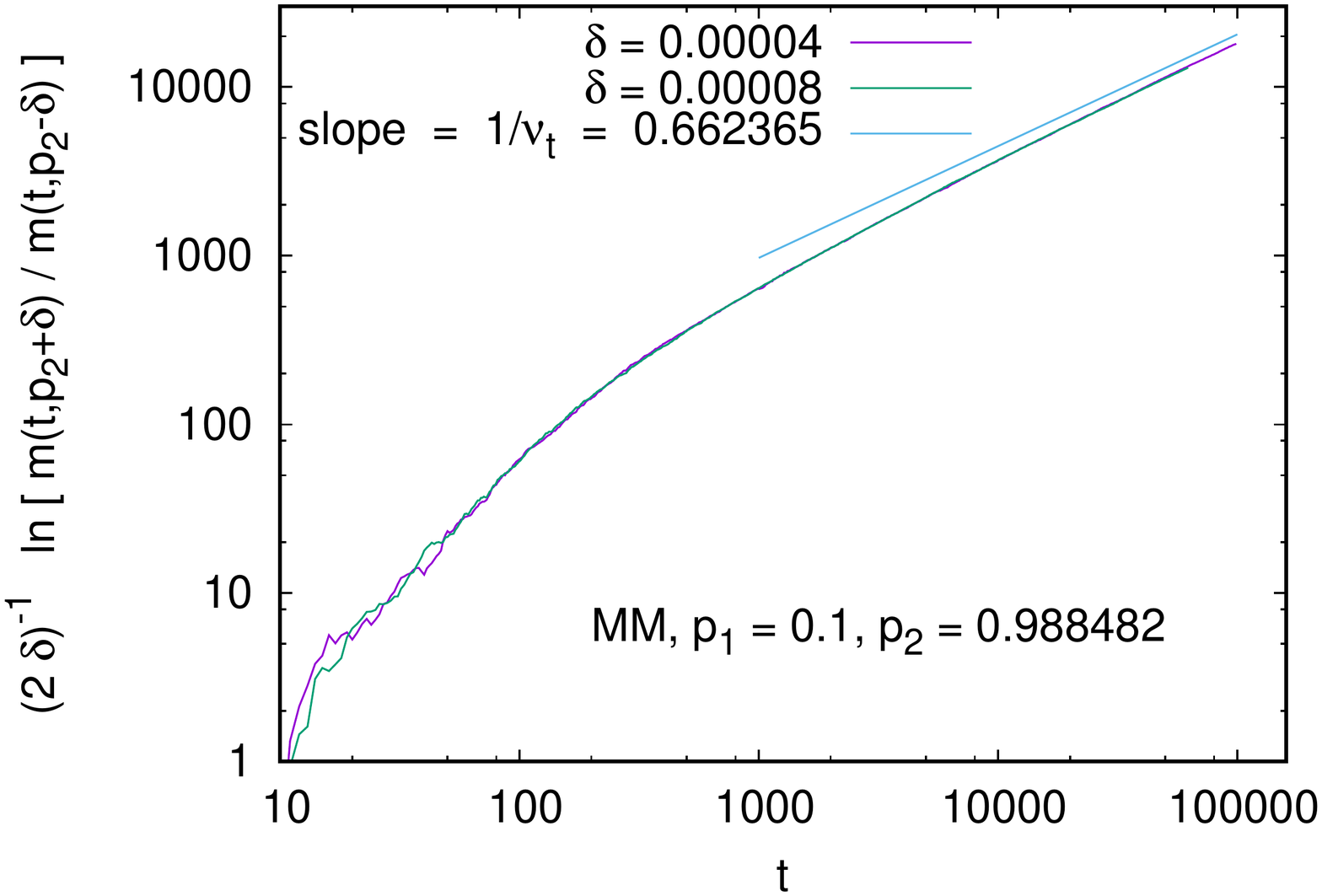}
\vglue -0.4cm
\par\end{centering}
\caption{\label{fig4} (color online) Log-log plot of the derivative of $\ln m(t,p_2)$ with respect to 
   $p_2$, plotted against $t$. Two estimates of the derivative are provided by finite difference quotients with 
   $\Delta p_2 = 0.00004$ and $0.00008$, respectively. The straight line is the scaling expected for OP,
with $\nu = 4/3$.}
\end{figure}

\begin{figure}
\begin{centering}
\vglue -0.4cm
\includegraphics[scale=0.30]{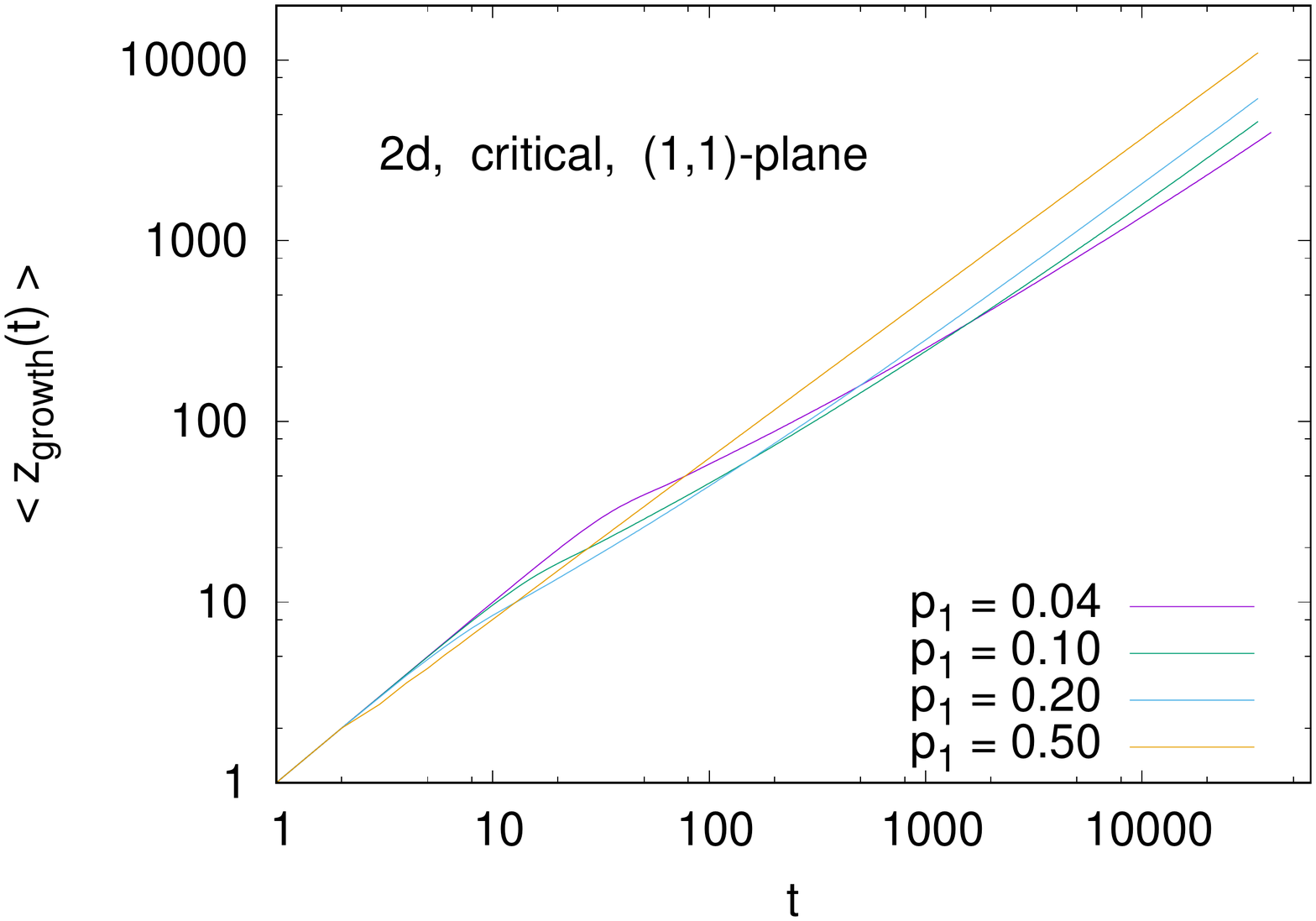}
\vglue -0.4cm
\par\end{centering}
\caption{\label{fig5} (color online) Log-log plot of $h(t)$ versus $t$ for critical interfaces of the 
   MM with initial orientation (1,1).}
\end{figure}

It was thus easy to verify that the model has for all values of $p_1$ the percolation critical exponents 
when it is {\it exactly on} the critical line. But verifying the exponents that control the off-critical 
behavior (the order parameter exponent $\beta$ and the correlation length exponent $\nu$ turned out 
to be much more difficult. One reason is the smallness of the critical region when $p_1$ is small.
Another reason is that corrections to scaling seem to be even more important. 

One popular way to obtain $\nu$ or $\nu_t = \nu d_{\rm min}$ is by means of data collapse plots. In 
the present case we would plot $t^{-\mu} m(t)$ against $(p_2-p_{2,c})t^x$ for different values of $x$, 
and would expect a data collapse when $x=1/\nu_t$. Due to the strong scaling violations for small 
$t$, one would do this only for values of $t$ that seem to be in the scaling region. For $p_1 = 0.1$,
Fig.~S1 would suggest that a good collapse can be expected for $t>10^4$. Unfortunately, when this is 
done, the best estimate for $\nu_t$ is about 10 percent smaller than the value for percolation.
The reason for this can be seen when plotting 
\be
   \frac{d\ln m(t,p_2)}{dp_2}|_{p_2 = p_{2,c}} \approx \frac{\ln [m(t,p_2+\epsilon)/m(t,p_2-\epsilon)]}{2\epsilon}
\ee
against $t$. From percolation theory we expect that this should scale as $t^{1/\nu_t}$. As seen from 
Fig.~4, this is compatible with the data, but these data do not fall strictly on a straight line 
even for the largest values of $t$. There is some visible curvature even for $t>10^4$, and the critical
scaling would be observed only for $t>10^5$. In spite of this difficulty, we believe that our data
are fully compatible with the OP universality hypothesis.  

\section{Line seeds for the MM}

If we want to study originally flat interfaces, we have to use line seeds. For small $p$ and $t$, the 
growth depends strongly on the orientation of the seed. For interfaces parallel to one of the 
coordinate axes, initial growth is extremely slow because its velocity is $\propto p$: The interface
can only progress, if a site with a single infected neighbor gets infected. On the other hand, for 
surfaces in the $(1,1)$ orientation (diagonal), growth is $\propto q$ and thus maximal. For $p\to 0$,
the original speed of growth is 1, and stays $\approx 1$ until a finite part of the interface lags
because $q<1$.

\begin{figure}
\begin{centering}
\includegraphics[scale=0.30]{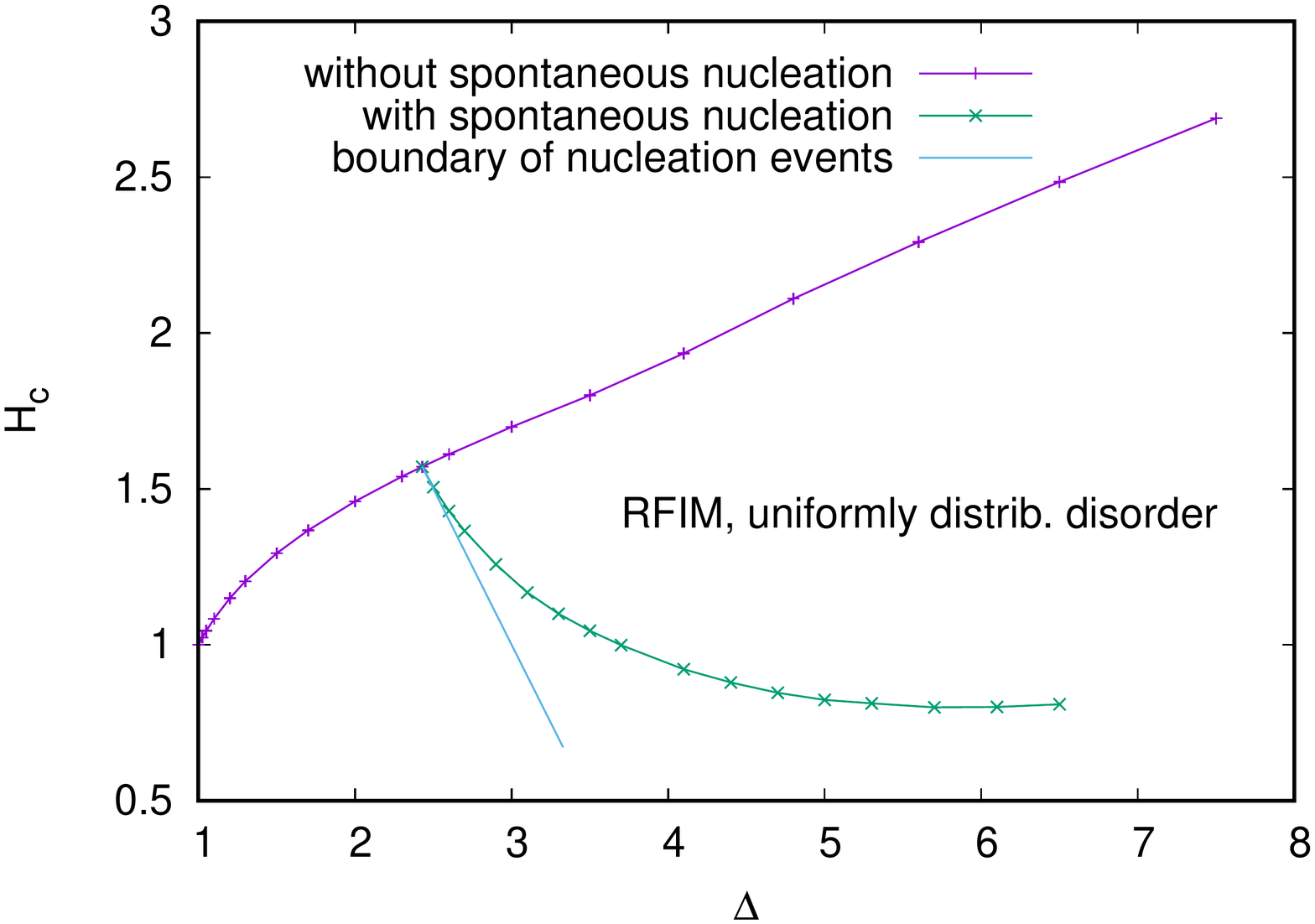}
\par\end{centering}
\caption{\label{fig6} (color online) Depinning thresholds for the RFIM with uniformly distributed 
   disorder. The value $H_c=1$ for $\Delta= 0$ is an exact analytic result, the other points are 
   from simulations and have error bars much smaller than the symbols. The straight line indicates 
   the boundary $H+\Delta=4$, below which no single spins can flip spontaneously.}
\end{figure}

\begin{figure}
\begin{centering}
\vglue -0.4cm
\includegraphics[scale=0.30]{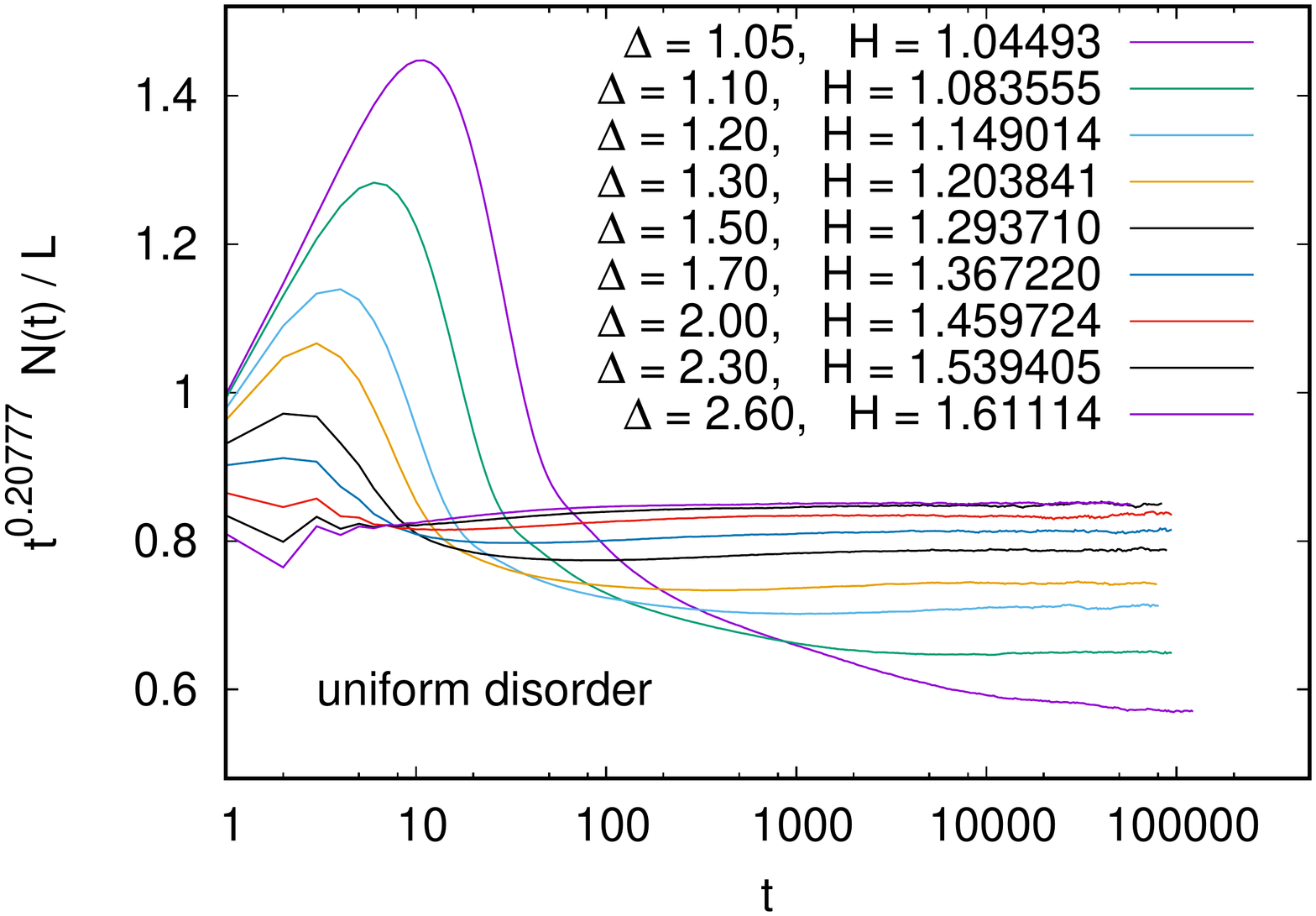}
\vglue -0.4cm
\par\end{centering}
\caption{\label{fig7} (color online) Log-linear plot of $t^{0.20777}N(t)/L$ versus $t$ for critical 
   interfaces with initial orientation (1,1), in the RFIM without spontaneous nucleation and with 
   uniform disorder.}
\end{figure}

Average heights of interfaces with $(1,1)$ orientation at criticality are shown in Fig.~S5, where
we defined heights via the heights of growth sites, $h(t) = \langle z_{\rm growth}(t)\rangle$.
We see that indeed $h(t) \approx t$ for small $p$ and $t < 1/p$, but for large $t$ all curves show
the expected $h(t)\sim t^{1/d_{\rm min}}$. 

\section{Line seeds for the RFIM}

The depinning thresholds for the RFIM with uniformly distributed disorder are shown in Fig.~S6.
As in the analogous Fig.~8 (main text) for the Gaussian model, the upper curve is for the version
where spontaneous spin flips are forbidden, while the lower curve is for the version where single
spins can flip. Such flips of isolated spins can occur only, if $H+\Delta >4$, which is indicated 
by the straight line in Fig.~S6. Therefore the depinning threshold curve for the version with single
spin flips must be above this line. It seems to be tangential for it at the point $\Delta = 2.42897(3),
H=4-\Delta$. The existence of a sharp disorder threshold below which no spins can flip spontaneously
is the main qualitative difference with the Gaussian model. Otherwise the curves are very similar. 
In particular, the depinning threshold curves for the versions with spontaneous spin flips are allowed
are non-monotonic with disorder. 

We verified that along both curves in Fig.~8 the exponents are those of critical percolation. This
even includes the point $\Delta = 2.42897(3),H=4-\Delta$, as far as the exponents $\delta,\mu,\eta$,
and $d_{\rm min}$ are concerned. The exponents $\beta$ and $\nu$ describing off-critical behavior 
for the version with spontaneous flips are of course different at this point, but they are also as
in OP everywhere else.

Finally, we should point out that for uniformly distributed 
disorder the thresholds depend rather strongly on the sizes of the clusters that are allowed to flip
spontaneously. If the dynamics is such that also pairs of adjacent spins can flip, then such flips 
can flip already for $H+\Delta >2$, while for any $H+\Delta >$ the probabilities for for flipping
large enough clusters are non-zero. 

We show here only some of the many data that verify the critical exponents -- both for the Gaussian 
and for the uniform model, and both for with and without single spin flips --, because they are so 
similar to those for the MM.
In Fig.~S7 we show, e.g., values of $N(t)/L$ for uniform disorder, at critical values of $H$. In 
order to achieve a high resolution on the y-axis, these data are multiplied by $t^{0.20777}$, so that OP 
scaling would give horizontal curves for large $t$. We see this is indeed the case, although
finite-$t$ corrections are huge for small disorder.

\section{The uniform noise RFIM with $\Delta < 1$}

For the RFIM with uniform noise and noise strengh $\Delta$, critically pinned interfaces are in the 
OP universality class for $\Delta > 1$. As $\Delta \to 1$ from above, also $H_c \to 1$. For $\Delta 
<1$, the flipping probability $p_1$ is non-zero only for $H>2-\Delta >1$. For $1 \leq H < 2-\Delta$, 
an interface parallel to one of the coordinate axes cannot grow, but since $p_2\geq 1$ (see Eq.~(S2)) any
tilted interface can grow with a finite speed that depends on the tilt angle (for (1,1) interfaces 
the speed is maximal). Finally, the cluster resulting from any finite seed configuration cannot
extend out of the convex hull of the seed, since that would require at least one spin with a single 
flipped neighbor to flip. If, however, the cluster growth is not done in an ``epidemic" paradigm,
but is done as in invasion percolation \cite{Wilkinson}, the growth of an infinite cluster is enforced
and its shape will be a convex polygon, i.e. the interface is facetted. In \cite{Qin-2012,Si-2016} 
it is claimed that the growth is ``first order" (i.e. discontinuous) for $\Delta <1$, which is slightly 
misleading. It is indeed discontinuous for tilted interfaces, but it is continuous for (0,1) interfaces.
Indeed, for any originally flat interface there are no overhangs, and the model actually is a simple
solid-on-solid model.